\documentclass[prb,twocolumn,showpacs,amsmath,amssymb,preprintnumbers]{revtex4}
\usepackage{bm}
\usepackage{graphicx}
\usepackage{color}

\begin{document}

\title{Magnetism and electrical transport in Y-doped layered iridate Sr$_2$IrO$_4$}

\author{Imtiaz Noor Bhatti}\affiliation{School of Physical Sciences, Jawaharlal Nehru University, New Delhi - 110067, India.}
\author{A. K. Pramanik}\email{akpramanik@mail.jnu.ac.in}\affiliation{School of Physical Sciences, Jawaharlal Nehru University, New Delhi - 110067, India.}

\begin{abstract}
Here, we report an investigation of structural, magnetic and electronic properties in Y-doped layered iridate (Sr$_{1-x}$Y$_x$)$_2$IrO$_4$ ($x$ $\leq$ 0.1). The parent Sr$_2$IrO$_4$ is a well-studied spin-orbit coupling (SOC) induced insulator with an antiferromagnetic ground state. The Y-doping here equivalently acts for electron doping without altering the vital parameters such as, SOC and electron correlation. Experimental results show a minute change in structural parameters and an equivalent charge conversion from Ir$^{4+}$ to Ir$^{3+}$. Unlike similarly other electron-doped system, the low temperature magnetic and electronic state in present series is minimally influenced. The charge conduction mechanism follows 2-dimensional hopping model in whole series. Magnetoresistance (MR) data show an interesting sign change with both temperature and magnetic field. The positive MR both at low temperature follows weak antilocalization behavior where the sign change in MR is believed to be caused by an interplay between SOC and magnetic moment.
\end{abstract}


\maketitle
\section{Introduction}
The relativistic spin-orbit coupling (SOC) is believed to drive novel ground state in 5$d$-based Sr$_2$IrO$_4$.\cite{krempa, rau,cao,caobook} This material has layered structure, and recently has received large deal of attention due to its unusual insulating and magnetic states. The structural similarity with high $T_c$ superconducting material i.e., doped La$_2$CuO$_4$ has further added interest in Sr$_2$IrO$_4$ where the band calculations have predicted possible superconductivity in this material with optimum electron/hole doping.\cite{yang,gao,casa,yan,sumita} The novel physics in Sr$_2$IrO$_4$ has been explained with $J_{eff}$ = 1/2 ground state.\cite{kim1,kim2} In octahedral environment, high crystal field splits the $d$ orbitals into $t_{2g}$ and $e_g$ levels. In presence of strong SOC, the low lying $t_{2g}$ state is further split into into $J_{eff}$ = 3/2 quartet and $J_{eff}$ = 1/2 doublet. Under this picture, Ir$^{4+}$ (5$d^5$) in Sr$_2$IrO$_4$ adopts a half filled $J_{eff}$ = 1/2 state. It has been further argued that even in presence of small electronic correlation ($U$), this narrow $J_{eff}$ = 1/2 state is split into upper (empty) and lower (filled) band which gives a realization of Mott-like insulating state. 
 
The structure has critical role in Sr$_2$IrO$_4$. The layered structure in this material is realized from generalized Ruddlesden-Popper series (SrIrO$_3$)$_n$(SrO) with $n$ = 1, where the Ir$^{4+}$ atoms form a square lattice in each layer with $J_{eff}$ = 1/2 pseudospins. These pseudospins participate in Heisenberg-type antiferromagnetic (AFM) exchange interaction which exhibits sizable anisotropy along in-plane and out-plane direction.\cite{fujiyama} The rotation/distortion of IrO$_6$ octahedra further induces Dzyaloshinskii-Moriya (DM) type antisymmetric interaction which is believed to give rise weak ferromagnetic (FM) behavior with ordering temperature $\sim$ 225 K.\cite{crawford,ye,imtiaz1,imtiaz2} This picture, however, implies neighboring ionic states i.e., Ir$^{3+}$ and Ir$^{5+}$ to be nonmagnetic ($J_{eff}$ = 0) as those will have fully filled $J_{eff}$ = 3/2, 1/2 and $J_{eff}$ = 3/2 states, respectively. It is worthy to note here that this picture of SOC is described considering an ideal cubic environment of IrO$_6$ octahedra. In reality, however, many materials deviate from this ideal condition. Therefore, different picture of spin-orbit entanglement such as, mixing of $J_{eff}$ states has been proposed for many materials. The prominent examples include pyrochlore iridates (Eu,Sm)$_2$Ir$_2$O$_7$,\cite{sagayama,asih} hexagonal lattice systems (Li,Na)$_2$IrO$_3$,\cite{mkim,biffin} hyperkagome lattice material Na$_4$Ir$_3$O$_8$,\cite{takagi} chain compound Sr$_3$CuIrO$_6$,\cite{liu} etc. In recent times, many theoretical proposals have been put forward to address this new physics originating from an interplay between SOC and noncubic crystal field in iridates, therefore, the magnetic nature of Ir ions, as a whole, need to be investigated properly.\cite{liu,choong,bhattacharjee,georg}

Considering the anisotropic nature in Sr$_2$IrO$_4$, the approaches of chemical doping both at Sr- and Ir-site have been adopted to understand the exotic magnetic state in this material. In case of Ir-site doping as in Sr$_2$Ir$_{1-x}$$M_x$O$_4$ ($M$ = Rh, Ru, Tb, Cu), a complete suppressing of magnetic ordering and insulating state has been observed at different level of doping concentration.\cite{clancy,qi,calder,yuan,wang,imtiaz3} On the other hand, the isoelectronic doping at Sr-site in (Sr,Ca,Ba)$_2$IrO$_4$ has shown a drastic decrease in resistivity while having insignificant influence on magnetic state.\cite{chen1,zhao,souri} The La$^{3+}$ in (Sr$_{1-x}$La$_x$)$_2$IrO$_4$, interestingly, has shown rapid suppression of both magnetic and insulating state with only $\sim$ 2\% of doping concentration.\cite{chen,gretarsson} Similarly, Ga$^{3+}$ doping in (Sr$_{1-x}$Ga$_x$)$_2$IrO$_4$ has shown a suppression of insulating state for $x$ $>$ 0.05 but the study shows magnetic transition temperature in Sr$_2$IrO$_4$ does not change significantly till $x$ = 0.15.\cite{han} As trivalent substitution at Sr-site converts Ir$^{4+}$ to Ir$^{3+}$ where the latter is expected to be nonmagnetic, it is surprising that robust magnetism in Sr$_2$IrO$_4$ with such high ordering temperature is suppressed with such small dilution of magnetic lattice.

In present work, we have undertaken Sr-site substitution with Y$^{3+}$ in (Sr$_{1-x}$Y$_x$)$_2$IrO$_4$. The Y$^{3+}$ (4$d^0$) being nonmagnetic, avoids any magnetic interaction with Ir at low temperature. Rather, it equivalently generates Ir$^{3+}$ (5$d^6$, $J_{eff}$ = 0) without causing any modification in crucial active interactions such as, SOC and $U$. Therefore, this substitution will act for electron-doping and will possibly introduce site-dilution through Ir$^{3+}$ in system. Here, it can be mentioned that both Y$^{3+}$ and La$^{3+}$ will introduce electron doping in system but the Y$^{3+}$ will have comparatively large impact on structural parameters due to larger ionic radii mismatch with Sr$^{2+}$. We have investigated structure, magnetization, electrical transport and magnetotransport in (Sr$_{1-x}$Y$_x$)$_2$IrO$_4$. The system retains its original crystal structure while there are nonmonotonic changes in structural parameters. X-ray photoemission spectroscopy (XPS) measurements suggest Y$^{3+}$ creates equal amount of Ir$^{3+}$. Both the magnetic and electronic states are minimally influenced with Y$^{3+}$ doping. While our analysis shows charge conduction mechanism follows Mott's two dimensional (D) variable-range-hopping (VRH) model, the magnetoresistance (MR) changes sign with both temperature as well as magnetic field which is due to an interplay between SOC and magnetism.

\section{Experimental Detail}
Series of polycrystalline materials (Sr$_{1-x}$Y$_x$)$_2$IrO$_4$ ($x$ = 0.0, 0.01, 0.02, 0.025, 0.035, 0.05 and 0.10) are prepared by conventional solid state method similar to Ref. [15]. The high purity powder ingredients SrCO$_3$, IrO$_2$ and Y$_2$O$_3$ (Sigma Aldrich, purity $\sim$ 99.99\%) are mixed in stoichiometric ratio and ground well. The Y$_2$O$_3$ has been given preheat treatment to remove moisture. The mixed powders have been given several heat treatment in temperature range (1000 - 1100)$^oC$ in powder and pellet form with intermediate grindings. X-ray diffraction (XRD) measurements have been used to check the phase purity of all the samples. The XRD data have been analyzed with Rietveld refinement program to get detail information of structure and phase purity. Further, x-ray photoemission spectroscopy (XPS) study has been done to probe the oxidation state of elements in parent and doped samples. The XPS measurements are performed with base pressure in the range of 10$^{-10}$ mbar using a commercial electron energy analyzer (Omnicron nanotechnology) and non-monochromatic Al$K\alpha$ x-ray source ($h\nu$ = 1486.6 eV). CasaXPS software has been used to analyze the XPS data. The samples used for XPS study are in pallet form where the ion beam sputtering has been done on samples to expose the clean surface before measurements. The DC magnetization have been measured using a  SQUID magnetometer (Quantum Design) whereas the electrical transport measurements are done in an integrated system from NanoMagnetics Instruments and Cryomagnetics, Inc.

\begin{figure}
	\centering
		\includegraphics[width=8cm]{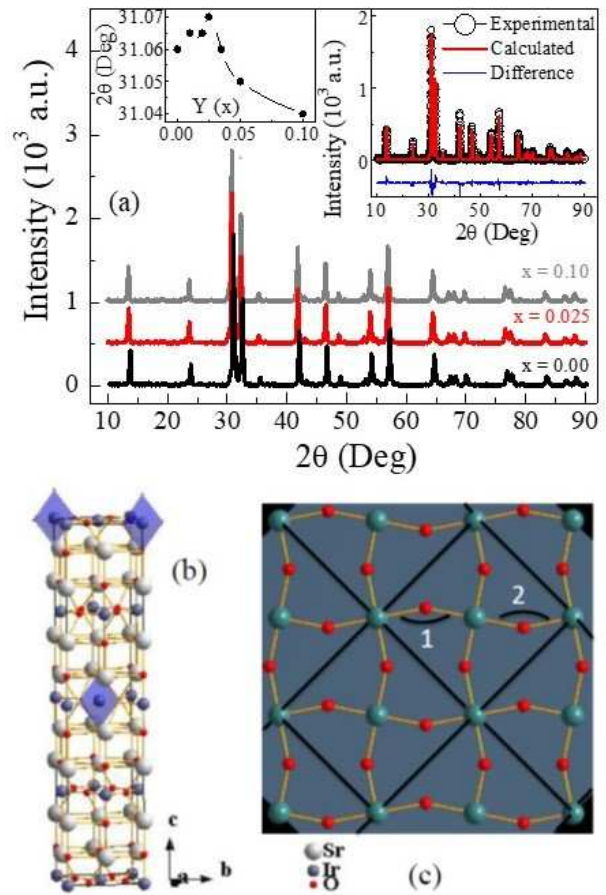}
	\caption{(color online) (a) The x-ray diffraction pattern are shown for three representative samples ($x$ = 0.0, 0.025 and 0.1) of (Sr$_{1-x}$Y$_x$)$_2$IrO$_4$ series. Left inset shows the position of most intense peak (2$\theta$ $\sim$ 31$^o$) as a function of Y($x$) doping, the line is a guide to eye. The right inset shows the XRD pattern along with Rietveld analysis for $x$ = 0.0 sample. (b) shows the structural unit cell for Sr$_2$IrO$_4$ and (c) depicts the Ir-O layer showing Ir square lattice. The 1 and 2 in figure mark the Ir-O-Ir bond angle.}
	\label{fig:Fig1}
\end{figure}

\section{Result and Discussion}
\subsection{Structural characterization}
The Fig. 1a shows room temperature XRD pattern of three representative samples with $x$ = 0.0, 0.025 and 0.1 of present series. The XRD data indicate there is no substantial changes in pattern with Y substitution which is important considering that Sr$^{2+}$ (1.21 \AA) and Y$^{3+}$ (0.96 \AA) has considerable differences in ionic radii (for coordination number = 7). The left inset of Fig. 1a shows composition dependent peak position of most intense peak situated at 2$\theta$ $\sim$ 31$^o$. The plot shows the maximum peak shifting is about 0.03$^o$ which is though nonmonotonous. Interestingly, the peak shifting displays an anomaly around $x$ = 0.025 which has been similarly observed for other lattice parameters (discussed later in Fig. 2). The structural analysis of materials has been done by Rietveld refinement of XRD data using FULLPROF program.\cite{full} Right inset shows the representative Rietveld refinement for parent $x$ = 0 material. Rietveld refinement of XRD data shows all the samples are in single phase and crystallize in tetragonal structure with \textit{I4$_1$/acd} symmetry.\cite{imtiaz1,imtiaz2} We have obtained $\chi^2$ for Rietveld refinement below $\sim$ 2 for all the samples which implies a good quality fitting. The tetragonal unit cell, as drawn using structural parameters obtained from Rietveld refinement, is shown in Fig. 1b for Sr$_2$IrO$_4$. It can clearly be seen the layered structure of present material where Ir-O active layers are separated by inactive Sr-O layers. Fig. 1c presents the Ir-O layer showing Ir square lattice where 1 and 2 in figure marks the Ir-O-Ir bond-angle.

While Y substitution do not induce any structural phase modification, the lattice parameters show a slight modification with $x$. Figs. 2a and 2b present evolution of lattice parameters $a$ and $c$ with Y concentration $x$ demonstrating both the parameters initially decrease slightly till $x$ $\sim$ 0.025 then increases rapidly with $x$. Similar anomalous changes of lattice parameter, which is related to interesting electronic properties, have been evidenced in doped pyrochlore iridate Eu$_2$Ir$_2$O$_7$.\cite{surjeet}

The overall change/increase in lattice parameters ($a$ and $c$) across the series is not very significant (below $\sim$ 0.2\%) where we obtain an overall increase in unit cell volume is $\sim$ 0.48\%. Although, both $c$ and $a$ parameters show similar change with $x$ (Figs. 2a and 2c), but the $c/a$ ratio decreases which implies tetragonal distortion is reduced with progressive substitution of Y. Fig. 2c presents variation of basal plane Ir-O bond angle $<$Ir-O1-Ir$>$ which shows similar dip around $x$ $\sim$ 0.05. This $<$Ir-O1-Ir$>$ is directly related to IrO$_6$ octahedral rotation ($\theta_{oct}$) around $c$-axis which plays a significant role in Sr$_2$IrO$_4$ as this octahedral distortion is believed to induce Dzyaloshinsky-Moriya (DM) type antisymmetric exchange interaction giving rise to canted AFM spin structure or weak ferromagnetism in this material. For parent Sr$_2$IrO$_4$, we find $\theta_{oct}$ around 11.3$^o$ deg  which is consistent with other reports (Fig. 2d).\cite{crawford, imtiaz1} With doping, $\theta_{oct}$ initially increases to 13.02$^o$ at $x$ = 0.025, and then with further Y substitution it decreases. The changes in structural parameters in Fig. 2 show a reverse trend around $x$ $\sim$ 0.025. This may be related to the fact that at lower concentration of Y$^{3+}$, which has smaller size compared to Sr$^{2+}$, both the cell parameters and volume decrease. However, the Ir$^{3+}$ (0.68 \AA), which is generated with Y$^{3+}$ substitution and has larger size compared to Ir$^{4+}$ (0.62 \AA), will increase the unit cell volume at higher concentration of Y$^{3+}$. Nonetheless, the observed variation in lattice parameters, bond angle $<$Ir-O1-Ir$>$ and octahedral distortion $\theta_{oct}$ are consistent with each other (Fig. 2).            

\begin{figure}
	\centering
		\includegraphics[width=8cm]{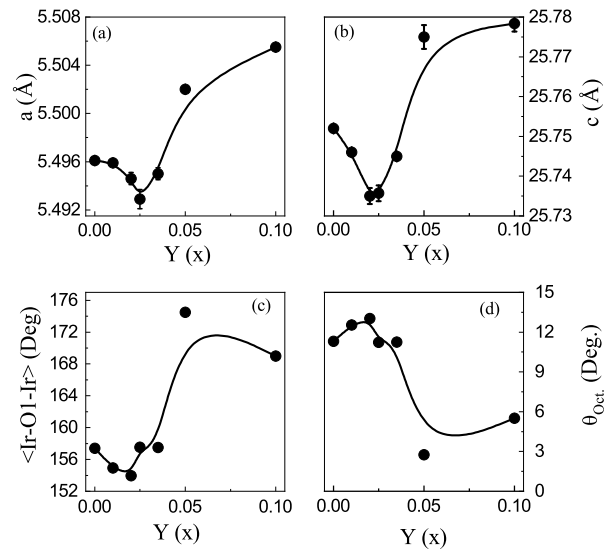}
	\caption{(color online) (a) shows lattice parameter $a$, (b) shows lattice parameter $c$ (c) shows bond angle $<$Ir-O1-Ir$>$ and (d) shows octahedral rotation $\theta_{Oct}$ as a function of Y concentration $x$ for (Sr$_{1-x}$Y$_x$)$_2$IrO$_4$ series. These parameters are obtained from Rietveld analysis of powder x-ray diffraction data.}
	\label{fig:Fig2}
\end{figure}

\subsection{X-ray photoemission spectroscopy study }
It is important to understand the electronic charge state of Ir in Sr$_2$IrO$_4$ and its evolution with Y substitution which decides the physical properties in these materials. With this aim, we have performed the XPS measurements for representative three samples i.e., $x$ = 0.0, 0.025 and 0.05. Figs. 3a, 3b and 3c show the Ir-4\textit{f} core level spectra for $x$ = 0.0, 0.025 and 0.05, respectively. The Ir-4\textit{f} spectra shows two distinct peaks which correspond to spin-orbit split Ir-4\textit{f}$_{7/2}$ and Ir-4\textit{f}$_{5/2}$ electronic states that are observed at the binding energies (B.E.) of 62.0 and 65.0 eV, respectively.\cite{zhu,kumar} The red lines in Figs. 3a, 3b and 3c are due to overall fitting of data. For $x$ = 0.0, the detailed analysis of Ir-4\textit{f} fitting indicates Ir ions are mostly in Ir$^{4+}$ charge states (blue solid lines). However, a weak contribution from neighboring ionic states (Ir$^{5+}$ or Ir$^{3+}$) can be present in parent Sr$_2$IrO$_4$ due to non-stoichiometry of samples which has been reported in other studies.\cite{imtiaz3,kumar} However, that feature is not clearly evident in present data. For the doped samples as in Figs. 3b and 3c, the similar data analysis indicates that in addition to Ir$^{4+}$, we find trace of Ir$^{3+}$ with Ir-4\textit{f}$_{7/2}$ and Ir-4\textit{f}$_{5/2}$ peaks arising at binding energies of 61.1 and 64.2 eV, respectively (orange solid lines). The origin of Ir$^{3+}$ in doped samples is related to Y substitution. Interestingly, we find that the amount of Ir$^{3+}$ increases almost proportionately with Y doping and its amount corresponds to expectations for the level of Y doing ($x$). For instance, we find amount of Ir$^{3+}$ is about 4.8\% and 9\% for $x$ = 0.025 and 0.05 samples, respectively. 

\begin{figure*}
	\centering
		\includegraphics[width=16cm]{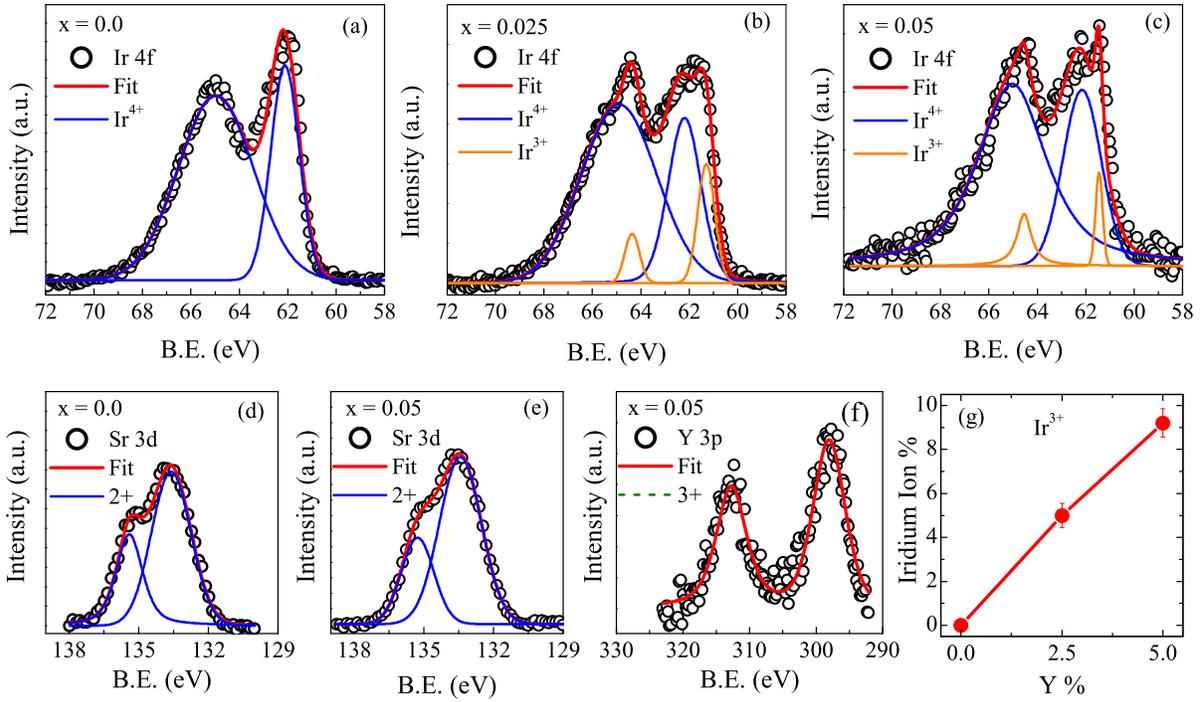}
	\caption{(color online) (a), (b) and (c) show the Ir-4$f$ core level spectra as a function of binding energy (B.E.) of $x$ = 0.0, 0.025 and 0.05, respectively for (Sr$_{1-x}$Y$_x$)$_2$IrO$_4$ series. Open black circles represent the experimental data and the red solid lines in (a), (b) and (c) are the fitted envelope taking contributions of Ir$^{4+}$ component for $x$ = 0.0 and Ir$^{4+}$ and Ir$^{3+}$ components for $x$ = 0.025 and 0.05 samples. The individual contribution from Ir$^{4+}$ and Ir$^{3+}$ components plotted as solid lines in blue and orange colors, respectively. (d) and (e) show the Sr-3$d$ core-level spectrum of $x$ = 0.0 and 0.05 samples, respectively. (f) shows the Y-3$p$ core-level spectrum of $x$ = 0.05 sample. (g) shows variation of Ir$^{3+}$ ions with Y concentration.} 
	\label{fig:Fig3}
\end{figure*}

The Figs. 3d and 3e show core-level spectra of the Sr-3\textit{d} orbitals for representation x = 0.0 and 0.05 samples, respectively where the spin-orbit coupling split doublet peaks related to Sr-3$d_{5/2}$ and Sr-$d_{3/2}$ are located at 133.3 and 135.1 eV, respectively. While the red continuous lines in Figs. 3d and 3e are the overall fitting and the solid blue lines are due to fitting with Sr$^{2+}$ electronic states. Similarly Fig. 3f shows XPS spectra of Y-3\textit{p}$_{3/2}$ core level for doped $x$ = 0.05 sample. We observe that Y-3$p_{3/2}$ and Y-3$p_{1/2}$ peaks related to Y$^{3+}$ occur at binding energy around 299.1 and 312 eV respectively with spin-orbit splitting of $\sim$ 12.9 eV. Confirming the Y$^{3+}$ charge state of Y is important which equivalently converts Ir$^{4+}$ to Ir$^{3+}$ and acts for electron doping in system. Fig. 3g shows the variation of Ir$^{3+}$ with Y$^{3+}$ doping concentration. Considering that Y($x$) generates 2$x$ amount of Ir$^{3+}$, it is understood that Y$^{3+}$ doping generates an equivalent amount of Ir$^{3+}$ ion. This will help to understand the role of Ir$^{3+}$ ions on magnetic and transport properties as within the picture of strong SOC, Ir$^{3+}$ electronic state is believed to be nonmagnetic.

\subsection{Magnetization study}
Temperature dependent magnetization data measured in 10 kOe magnetic field under zero-field cooled (ZFC) and field cooled (FC) protocol are shown in Figs. 4a ($x$ = 0.0, 0.01 and 0.02) and 4b ($x$ = 0.025, 0.035, 0.05 and 0.1) for (Sr$_{1-x}$Y$_x$)$_2$IrO$_4$ series. For all the samples, $M(T)$ data show a sharp increase below around 250 K which is marked by magnetic phase transition. In Sr$_2$IrO$_4$, a magnetic transition from high temperature paramagnetic (PM) to canted-type AFM state at low temperature has been identified.\cite{crawford, imtiaz1} This canted type AFM spin ordering originates due to DM interaction which is induced by distortion/rotation of IrO$_6$ octahedra. As a result, the system exhibits weak ferromagnetism. Using a critical behavior analysis, we have recently shown the nature of FM state in Sr$_2$IrO$_4$ closely follows the mean-field model with FM ordering temperature $T_c$ $\sim$ 225 K which agrees also with other studies.\cite{fujiyama,imtiaz2} The Fig. 4a further shows a downfall in ZFC $M(T)$ data and opening of bifurcation between $M_{ZFC}$ and $M_{FC}$ below $\sim$ 95 K. This downfall in $M(T)$ has been shown to arise due to prominent magneto-structural coupling where the spin alignment follows structural evolution at low temperature.\cite{imtiaz1,jack}  

\begin{figure}[t]
	\centering
		\includegraphics[width=8cm]{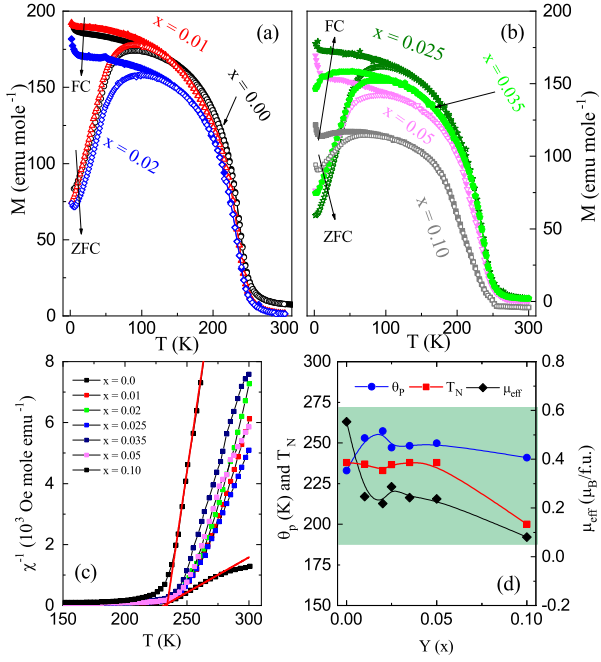}
	\caption{(a) and (b) show the temperature dependence of magnetization data measured in applied field of 10 kOe following ZFC and FC protocol for (Sr$_{1-x}$Y$_x$)$_2$IrO$_4$ series. Open and filled symbols represent ZFC and FC data, respectively. (c) shows temperature dependent inverse susceptibility ($\chi^{-1}$ = $(M/H)^{-1}$) as deduced from ZFC magnetization data that are shown in (a) and (b). The solid lines are due to fitting with Curie-Weiss law for representative two samples $x$ = 0.0 and 0.1. (d)  Variation of $\theta_P$, $T_N$ and $\mu_{eff}$ are shown as a function of $x$ for (Sr$_{1-x}$Y$_x$)$_2$IrO$_4$ series.}
	\label{fig:Fig4}
\end{figure}

It is evident in Figs. 4a and 4b that Y substitution has very weak influence on magnetization as the AFM transition temperature $T_N$, magnetic moment and bifurcation between magnetization data do not change significantly till $x$ = 0.05. For $x$ = 0.1, the figure shows a slight decrease in $T_N$ as well as in magnetic moment. This is significant as unlike other electron-doped system (Sr$_{1-x}$La$_x$)$_2$IrO$_4$ where the magnetic state collapses with $x$ $\sim$ 0.02, the magnetic state in present (Sr$_{1-x}$Y$_x$)$_2$IrO$_4$ is minimally disturbed. Given that both Y$^{3+}$ and La$^{3+}$ generates Ir$^{3+}$ and dilutes the magnetic lattice at same scale, the robust magnetic behavior in present series is quite surprising. Here, it can be further added that in another electron-doped (Sr$_{1-x}$Ga$_x$)$_2$IrO$_4$, though the order magnetic moment decreases but the magnetic transition temperature remains mostly unchanged with $x$ or Ir$^{3+}$ even till $x$ = 0.075.\cite{han} This indicates that $J_{eff}$ picture of Ir under non-cubic octahedral distortion needs to be reexamined.      

To characterize the effective moment in the high temperature PM state, temperature dependent inverse magnetic susceptibility $\chi^{-1}(T)$ are shown in Fig. 4c for this series. For $x$ = 0.0, $\chi^{-1}(T)$ shows linear behavior above $T_N$, however, in high temperature regime around 275 K, the $\chi^{-1}$(T) deviates from linearity. This deviation arises due to layered structure of Sr$_2$IrO$_4$ which gives an anisotropic magnetic exchange interaction where the strength of exchange interaction is different within the layer and between the layers. A scattering study has shown 2D like anisotropic magnetic interaction in Sr$_2$IrO$_4$ which allows in-plane magnetic interaction to survive at least till $\sim$ 25 K above the $T_N$ (228.5 K).\cite{fujiyama} Therefore, the slope change in $\chi^{-1}(T)$ for Sr$_2$IrO$_4$ around 275 K (Fig. 4c) is likely due to anisotropic magnetic exchange coupling. The continuation of spin correlation above $T_N$ appears to be typical feature of layered AFM materials such as, La$_2$CuO$_4$,\cite{keimer} and iron pnictides.\cite{klingeler} In doped samples, as seen in Fig. 4c, the $\chi^{-1}$(T) becomes almost linear above $T_N$. The $\chi^{-1}$ data in PM regime is fitted with Curie-Weiss (CW) law, given as $\chi$ =$C/(T - \theta_P$) where $C$ (= N$_A$$\mu^2$$_{eff}$/3k$_B$) is the Curie constant, $\mu_{eff}$ is the effective PM moment and $\theta_P$ is the Curie temperature. Straight lines in Fig. 4c are due to fitting with CW law for two representative extreme members of series i.e., $x$ = 0.0 and $x$ = 0.1. Using the fitted parameter $C$, we have calculated the effective PM moment $\mu_{eff}$ for all the samples. 
 
\begin{figure}[t]
	\centerline
		{\includegraphics[width=8cm]{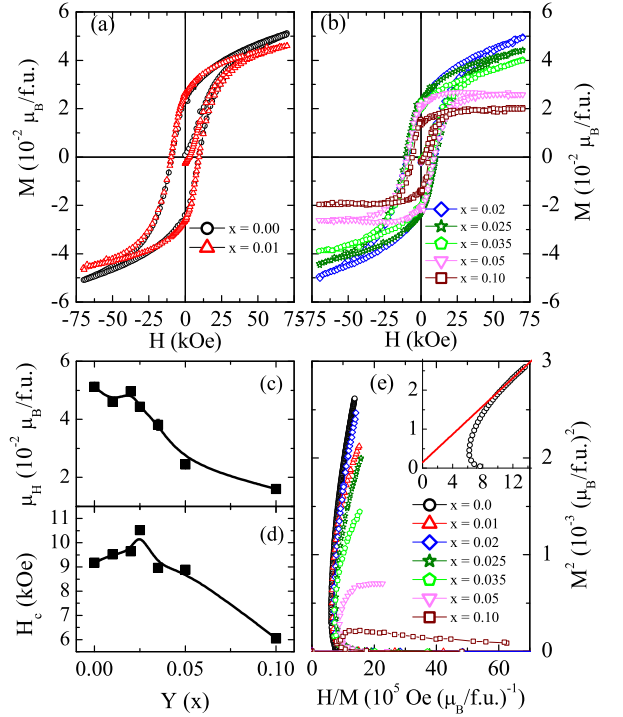}}
	\caption{((a) and (b) show the isothermal magnetization as function of magnetic field up to $\pm$ 70 kOe collected at 2 K for (Sr$_{1-x}$Y$_x$)$_2$IrO$_4$ series. (c) and (d) show the magnetic moment ($\mu_H$) at 70 kOe and the coercive field ($H_c$) with $x$, respectively. (e) shows the Arrott plots as obtained from $M(H)$ data for (Sr$_{1-x}$Y$_x$)$_2$IrO$_4$. Inset shows the magnified view of Arrott plot for $x$ = 0.0 material where the solid line is due to straight line fitting at high field regime.}
	\label{fig:Fig5}
\end{figure}

The Fig. 4d shows composition dependent $T_N$, $\mu_{eff}$ and $\theta_P$ of present series. We calculate $\mu_{eff}$ = 0.553 $\mu_B$/f.u. for $x$ = 0.0 composition which appears much lower than the calculated spin-only value 1.72 $\mu_B$/f.u. (= $g\sqrt{S(S+1)} \mu_B$) for spin-1/2 material. The $\mu_{eff}$ mostly decreases, though there are rapid decrease at initial level of $x$ and above $x$ = 0.02, the decrease of $\mu_{eff}$ becomes slower. Similar, though opposite, trend has been found for $\theta_P$ where it increases rapidly till $x$ = 0.02 then exhibits a slow change. The $\theta_P$ for parent Sr$_2$IrO$_4$ is as high as 233 K and the value is close to its long-range ordering temperature $T_N$ = 225 K.\cite{crawford,ye,imtiaz2} Nonetheless, changes in $\theta_P$ is very small over the series and it remains positive which is surprising because in highest doped sample i.e., $x$ = 0.1, there are conversion of $\sim$ 20\% Ir$^{4+}$ to Ir$^{3+}$ which would cause major dilution of magnetic lattice. From structural organization, Sr$_2$IrO$_4$ has layered structure where magnetically and electronically active Ir-O layers are separated by inactive Sr-O layer (Fig. 1b). For such 2-dimensional lattice, the geometrical percolation threshold for spin-spin only exchange interaction is shown to be around 40\% dilution.\cite{clancy} A recent study in doped Sr$_2$Ir$_{1-x}$Rh$_x$O$_4$ has shown complete disappearance of magnetism at $x$ $\sim$ 0.17.\cite{clancy} Taking into account Rh$^{3+}$ (4$d^6$) is itself nonmagnetic and it further generates nonmagnetic Ir$^{5+}$ ($J_{eff}$ = 0), this suggests around 34\% dilution. In that scenario, the maximum 20\% dilution in present (Sr$_{1-x}$Y$_x$)$_2$IrO$_4$ series should suppress the magnetic and insulating state to a large extent which is not evident in results. The La substitution, on the other hand, in (Sr$_{1-x}$La$_x$)$_2$IrO$_4$ has led to complete suppression of long-range magnetic and insulating state with only $x$ $\sim$ 0.02,\cite{chen,gretarsson} where the Ir ionic state would be similar in both La- and Y-doped materials. In another electron doped system (Sr$_{1-x}$Ga$_x$)$_2$IrO$_4$, though a metallic state is induced in Sr$_2$IrO$_4$ for $x$ $>$ 0.05 but the long-range transition temperature does not charge much with introduction of Ga$^{3+}$ (till $x$ = 0.075) in system.\cite{han} In an ideal and cubic octahedral environment, Ir$^{3+}$ (5$d^6$) is expected to be nonmagnetic with an impression that 5$d$ oxides possess high value of crystal field effect, hence all the electrons will populate low lying $t_{2g}$ state. However, iridates exhibit prominent SOC. Recent calculations and experimental results (RIXS), however, indicate a mixing of $e_g$ and $t_2g$ orbitals and $J_{eff}$ = 3/2 and $J_{eff}$ = 1/2 states due to prominent effects of SOC and non-cubic crystal field effect.\cite{liu,georg} In that case, Ir$^{3+}$ may not turn out to be totally nonmagnetic. Due to large mismatch between Sr$^{2+}$ and Y$^{3+}$ ionic size, such probability will be prominent in present (Sr1-xYx)2IrO4 series, the IrO$_6$ octahedra would be in non-ideal and non-cubic environment.  

Magnetic field dependent magnetization $M(H)$ measured at 2 K are shown in Figs. 5a and 5b for (Sr$_{1-x}$Y$_x$)$_2$IrO$_4$ series. The $M(H)$ for Sr$_2$IrO$_4$ shows a large opening at low temperature with coercive field $H_c$ $\sim$ 9370 Oe. With Y substitution, the $M(H)$ data show a saturation in high field regime. Moreover, the measured magnetic moment $\mu_H$ at 2 K and in field 70 kOe decreases with $x$ but shows an anomaly around $x$ = 0.025 (Fig. 5c). The $H_c$ similarly increases till $x$ = 0.025 then begins to decrease till highest doping level $x$ = 0.1 (Fig. 5d). Interestingly, this variation of $\mu_H$ and $H_c$ with $x$ has very similarity with $\mu_{eff}$ and $\theta_P$, respectively (see Fig. 4d). In Fig. 5e, we have shown Arrott plot of $M(H)$ data. Arrott plot ($M^2$ vs $H/M$) is an effective mean to understand the nature of magnetic state as the positive intercept on $M^2$ axis due to straight line fitting in high field regimes of Arrott plot implies spontaneous magnetization or FM state.\cite{arrott} For Sr$_2$IrO$_4$, Fig. 5e shows very small positive intercept (magnified view is in inset) which indicates weak ferromagnetism in this material. With Y substitution, we observe an increased positive intercept in Arrott plot which implies an increased spontaneous magnetization or FM behavior (Fig. 5e) in doped samples. This behavior, however, is consistent with $M(H)$ saturation (Fig. 5b) and increasing $\theta_P$ value (Fig. 4d) with doping concentration.

The above results show that magnetic properties are not significantly modified in present (Sr$_{1-x}$Y$_x$)$_2$IrO$_4$ series. In Sr$_2$IrO$_4$, various optical spectroscopy studies have shown results in favor of $J_{eff}$ model. In present series, a distortion in IrO$_6$ octahedra increases due to larger mismatch between Sr$^{2+}$ and Y$^{3+}$ ions. Following $J_{eff}$ model, Ir$^{3+}$ would be nonmagnetic and will act for site dilution. But our results show magnetic state is not modified significantly, thus implying Ir$^{3+}$ retains magnetism. Therefore, in non-ideal case of IrO$_6$ octahedra the theoretical investigations need to be extended to understand the effective $J_{eff}$ model. Recently, many theoretical studies are debating over the conventional $J_{eff}$ = 1/2 Mott-like insulating picture where this model is based on simplified local cubic symmetry of IrO$_6$ octahedra. For instance, the band topology in Na$_2$IrO$_3$ is explained with an improved model to clarify the topological character of spin-orbit coupled ground state.\cite{choong} In another report, the distortion of IrO$_6$ is shown to generate a new energy scale which competes with existing SOC and results in state of pseudospin that is significantly different from $J_{eff}$ = 3/2 and 1/2 multiplets.\cite{bhattacharjee} A recent detailed calculation has discussed the effect of crystal field effect, $U$ and SOC on mixing of $t_{2g}$ and $e_g$ orbitals, and its consequences on physical properties.\cite{georg} Nonetheless, our present study indicates a magnetic nature of Ir$^{3+}$ which probably gives robust magnetism in (Sr$_{1-x}$Y$_x$)$_2$IrO$_4$.

\subsection{Electronic transport study}
Electrical resistivity ($\rho$) as a function of temperature is shown in Figs. 6a and 6b for (Sr$_{1-x}$Y$_x$)$_2$IrO$_4$ series. The Sr$_2$IrO$_4$ shows strong insulating behavior which is believed to arise due to opening of exotic Mott-like gap in $J_{eff}$ = 1/2 electronic state.\cite{kim1} As evident in Fig. 6, all the samples remain highly insulating unlike its La-doped counterpart.\cite{chen} The changes of $\rho(x)$ at 2 K exhibit a nonmonotonic behavior showing a dip around $x$ = 0.025 (inset of Fig. 6b). Although the Y$^{3+}$ tunes the band filling proportionally through electron doping but it is not expected to alter both SOC and $U$, which unlike cases of Ir-site doping.\cite{clancy,calder,imtiaz3} This nonmonotonic changes of $\rho(x)$ is likely to be associated with modification of local crystal structures, such as Ir-O bond-angle and bond-length which show an anomalous changes across $x$ = 0.025 (see Fig. 2 ). The simultaneous decrease of both Ir-O bond-angle and bond-length will lead to further overlapping of Ir-$d$ and O-$p$ orbitals which would facilitate the electron hopping. As a consequence, the $\rho(x)$ initially decreases with $x$ and then again increases with opposite changes in structural parameters for $x$ $>$ 0.025 (inset of Fig. 6b).

\begin{figure}[t]
	\centerline
		{\includegraphics[width=8cm]{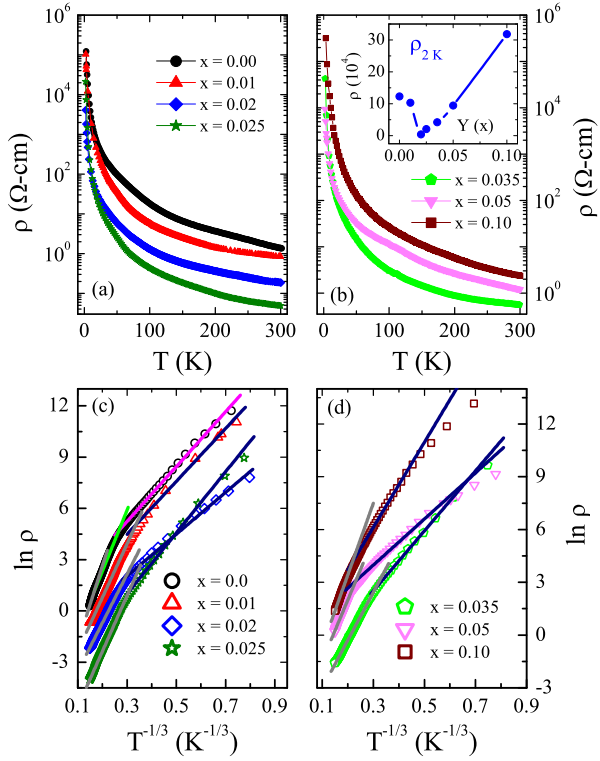}}
	\caption{(a) and (b) show the temperature dependent resistivity for (Sr$_{1-x}$Y$_x$)$_2$IrO$_4$ series. Inset shows the variation of resistivity at 2 K with composition $x$. (c) and (d) show the logarithm of resistivity as a function of $T^{-1/3}$ for (Sr$_{1-x}$Y$_x$)$_2$IrO$_4$. Solid lines are due to fitting with Mott's 2D VRH model in different temperature ranges.}
	\label{fig:Fig6}
\end{figure}

We find that this thermally activated charge conduction in present series can be best described by Mott's 2D VRH model, $\rho$ = $\rho_0$ $\exp\left(T_0/T\right)^{1/3}$, where $T_0$ = 21.2/[$k_B$$N(E_F$)$\xi^3$] is the characteristic temperature, $k_B$ is the Boltzmann constant, $N(E_F$) is the density of states (DOS) at Fermi level and $\xi$ is the localization length.\cite{mott} We have previously shown that charge conduction in Sr$_2$IrO$_4$ follows Mott's 2D VRH model in three distinct temperature regimes which are connected to its magnetic and structural state.\cite{imtiaz1,imtiaz3} For the doped samples, the $\rho(T)$ data can similarly be analyzed with VRH model but in two distinct temperature regimes. Figs. 6c and 6d show $\ln \rho$ vs $T^{-1/3}$ plotting where the straight line fitting confirms the validity of 2D VRH model. The temperature range of fitting as well as obtained $T_0$ parameter are given in Table 1. The parameter $T_0$ exhibits a nonmonotonic change where its value initially decreases and then increases similar to resistivity value $\rho (x)$ (inset of Fig. 6b). As $T_0$ is contributed by both DOS and localization length, therefore its variation with Y substitution can not be straightforwardly ascribed to any one parameter. Given that $\rho (x)$ shows nonmonotonic change and these materials are highly resistive, we speculate that change in $T_0$ is most likely related with change in $\xi$.

\begin{table}
\centering
\caption{\label{label} Temperature range and the parameter $T_0$ obtained from fitting of Mott's 2D VRH model with $\rho(T)$ data are given for (Sr$_{1-x}$Y$_x$)$_2$IrO$_4$ series.}
\begin{tabular}{ccc}
\hline
Sample &Temperature range (K) &$T_0$ (K)  \\
(Sr$_{1-x}$Y$_x$)$_2$IrO$_4$ &  &(10$^5$)\\
\hline
x = 0.0		&300 - 240  &1.441(3) \\
					&240 - 70 	&0.468(2) \\
					&40 - 5 		&0.048(1) \\
\hline
x = 0.01	&300 - 45 &0.307(8) \\
					&45 - 5  	&0.039(7) \\
\hline
x = 0.02	&300 - 40 &0.230(7)  \\
					&40 - 5 &0.018(3) \\
\hline
x = 0.025  &300 - 40 &0.345(2) \\
					&40- 5 	&0.025(7) \\
\hline
x = 0.035	&300 - 30 &0.202(7) \\
					&30 - 5  	&0.049(6) \\
\hline
x = 0.05	&300 - 110 &0.420(3)  \\
					&70 - 10 &0.022(4) \\
\hline
x = 0.10  &300 - 130 &0.676(2) \\
					&120- 10 	&0.141(7) \\
\hline
\end{tabular}
\end{table}

\subsection{Magnetoresistance}
The magnetoresistance (MR), calculated as $\Delta \rho/\rho(0)$ = $\left[\rho(H) - \rho(0)\right]/\rho(0)$, along with $M(H)$ data are shown in Figs. 7a - 7f for (Sr$_{1-x}$Y$_x$)$_2$IrO$_4$ series. The MR value is not very impressive (MR $\sim$ 2\%) but its evolution with temperature and magnetic field is interesting. For Sr$_2$IrO$_4$, MR at 2 K initially exhibits slight positive value but with increasing field it changes its sign to negative value (Fig. 7a). During decreasing magnetic field, MR notably shows large hysteresis showing the higher value compared to its value during increasing field. There is a remnant MR at $H$ = 0 which implies resistivity does not return to its original value at $H$ = 0 Oe. During application of negative magnetic field, MR initially shows a positive value indicating resistivity increases in presence of magnetic field. The MR then shows a peak and changes its sign in higher fields. On Negative side, with decreasing field the MR surprisingly do not show remnant value where it returns to its original value. With stark difference, MR at 300 K in nonmagnetic state shows complete positive value, though its values are almost two order lower than that at 2 K (Fig. 8b). For the doped samples, the MR shows similar hysteresis and sign change but it becomes more symmetric with magnetic field. For instance, positive MR increases during first application of magnetic field. Nonetheless, this behavior of MR appears to be connected with an interplay between SOC and magnetic moment. Further, it can be noted that the magnetic fields, where MR at 2 K shows peaks in both positive and negative field direction, closely match with coercive field $H_c$ in $M(H)$ data (Fig. 7).   

\begin{figure*}[t]
	\centerline
		{\includegraphics[width=16cm]{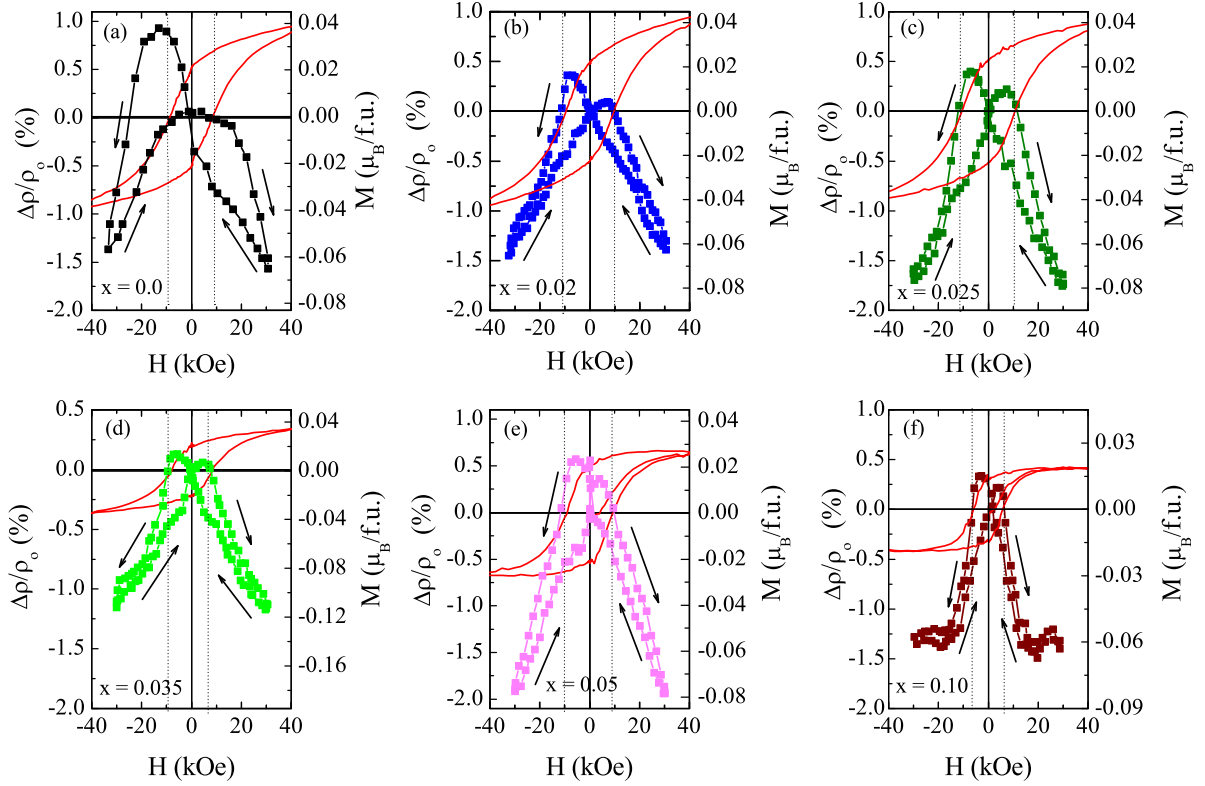}}
	\caption{(a), (b), (c), (d), (e) and (f) show the magnetoresistance (left axis) and the isothermal magnetization (right axis) as a function of magnetic field at 2 K. The vertical dotted lines mark the coercive field of respective samples.}
	\label{fig:Fig7}
\end{figure*}

The MR behavior in Fig. 7 has hysteresis though it shows an asymmetric behavior with respect to applied field. At a first glance, this MR appears to be driven by effect of `spin-polarized-tunneling' which has usually been observed in polycrystalline materials due to grain boundaries.\cite{coey,huang,gupta} The `spin-polarized-tunneling' induced MR basically has following properties; i) MR arises due to an increasing conductance at grain boundary which largely depends on the relative orientation of moment in adjacent grains, ii) This MR is mostly of negative type, as the applied magnetic field only polarizes the moments in adjacent grains which leads to an enhanced conduction, iii) This MR generally shows a high value, around 20-30\% or above, iv) As the spin polarization usually requires low field, this MR increases sharply in low field regime (below $\sim$ 0.5 T), and then increases at slower pace. The MR observed in present (Sr$_{1-x}$Y$_x$)$_2$IrO$_4$ series, on the other hand, has different characters. The observed MR in Fig. 7 is considerably low ($<$ 2\%), having both positive and negative value with the applied field. The MR here do not show any low/high field phenomenon. The positive MR is further asymmetric to the direction of applied field, and also sensitive to Y-doping or magnetic strength. The positive-negative crossover in MR roughly coincides with the coercive field $H_c$. Therefore, the present MR is unlikely due to `spin-polarized-tunneling' effect.

Generally, MR in disordered materials exhibits either positive or negative value which is explained using quantum interference phenomenon that is regarded as quantum correction to classical Drude model. While the negative MR is believed to originate from weak localization (WL), the positive MR arises due to weak antilocalization (WAL) effect in electronic transport mechanism. In zero magnetic field, the conductance is suppressed and enhanced in case of WL and WAL, respectively due to coherent back scattering. While traveling back to its original path with an opposite direction, an electron can follow many closed paths. Obviously, electrons have to follow those paths which only allow elastic scattering, otherwise electrons will loose their phase coherence that will eventually decrease the interference effect. Given that total amplitude of back scattering is the sum of amplitudes gained in both clockwise and anticlockwise direction, then in absence of magnetic field, magnetic impurities or any other dephasing scattering entities the calculation shows the amplitude to return to its original position is 2$\psi_0 \cos (\phi/2)$, where the amplitude $\psi_0$ is due to the phase accumulated in spatial part of the wave function and $\phi$ is the angle accumulated in spin part of the wave function.\cite{brahlek} In a system without spin-momentum locking, the $\phi$ part turns out to be significantly small, therefore the probability of backscattering in this constructive interference comes out as 4$|\psi_0|^2$. This enhanced backscattering reduces conductivity and causes WL effect. On the other hand, the prominent spin-momentum locking renders $\phi$ = $\pi$, thus this destructive interference gives almost zero backscattering which is the case for WAL. Now, in an applied magnetic field the time reversal symmetry is broken, and the two opposite paths acquires two equal but opposite phase $\beta$ = $e \Phi/\hbar$, where $\Phi$ is the related magnetic flux and $\hbar$ is reduced Planck constant. This additional phase factor $\beta$ modifies the previous amplitude to 2$\psi_0 \cos (\phi/2 + \beta)$. Therefore, with an increase of magnetic field ($\beta$ increases) the previously discussed constructive and destructive interference behaves differently. Nonetheless, even a small magnetic field contributes to the destruction of this localization process, hence showing a peak in magnetoconductance data near zero field. In materials with spin-momentum locking as induced by topological surface states or strong SOC, the WAL gives positive MR. The examples include Bi$_2$Se$_3$,\cite{chen2,liu} Bi$_2$Te$_3$,\cite{he} Au covered Mg film,\cite{berg} Ir-based Na$_2$IrO$_3$\cite{jender}, etc.

The Sr$_2$IrO$_4$ has reasonable SOC due to presence of heavy Ir. Previous studies have shown negative MR in bulk Sr$_2$IrO$_4$,\cite{chen,ge} however, a sign change in MR from positive to negative with temperature has also been observed in its film.\cite{miao} The hysteresis as well as sign change in MR in present study indicate MR is governed by an interplay between SOC and magnetism. At 300 K, where the system is in nonmagnetic state, a positive MR is observed. At low temperature magnetic state, a positive MR is observed at the beginning of field application which is again due to prominent SOC in present material. With increasing field, the magnetic moment increases which leads to breaking of time reversal symmetry, hence the moment acts as dephasing factor for electronic interference. The positive MR is, therefore, suppressed and the overall system exhibits negative MR which continues to increase till highest measuring field. The hysteresis and remnant MR is due to an induced moment with application of field. The exotic interplay between SOC and magnetism becomes more evident when MR shows a sign change from negative to positive value with reversal of magnetic field direction (Fig. 7). In an usual $M(H)$ loop, as magnetic field is reversed to negative direction, the remnant magnetic moment decreases and becomes zero around the coercive field $H_c$. Due to suppressed moment in this field regime between 0 to $H_c$, the SOC dominates and results in a positive MR. In the field higher than $H_c$, the moment further increases which promotes negative MR (Fig. 7). Interestingly, the peak position in positive MR closely matches with $H_c$.

In doped materials, Fig. 7 shows that the hysteresis in MR decreases with $x$ and the MR becomes more symmetric with field i.e., positive MR increases during first application of field. This increase of positive MR in doped samples is probably due to weakening of magnetism with Y substitution (Fig. 5c) which is also supported by reduced hysteresis in MR data. The influence of magnetic moment on MR appears very evident in highest doped $x$ = 0.1 sample where both MR and $M(H)$ exhibit saturation in high field regime (Fig. 7f). 

The effect of charge localization on magnetoconductivity (MC) $\Delta$$\sigma$ has been theoretically described by Hikami-Larkin-Nagaoka (HLN) equation.\cite{hln} This calculation basically neglects inelastic scattering compared to intrinsic elastic and spin-orbit scatterings and gives following form,

\begin{eqnarray}
\Delta\sigma = - \alpha \frac{e^2}{\pi h} \left[ln\left(\frac{\hbar}{4 e \textit{l}_{\phi}^2 B} \right) - \psi\left(\frac{1}{2} + \frac{\hbar}{4 e \textit{l}_{\phi}^2 B}\right)\right]
\end{eqnarray}

where $\Delta$$\sigma$ = $\sigma$(B) – $\sigma$(0), $e$ is the electronic charge, $h$ is Planck constant, $l_{\phi}$ is phase coherence length and $\alpha$ is prefactor which indicates type of localization i.e., its positive and negative value signifies WL and WAL, respectively. The value of $\alpha$ is predicted to be 0.5 for single channel WAL.

The MR at 2 K exhibits positive value and a dip like feature near origin in negative field side for Sr$_2$IrO$_4$ (see Fig. 7a). This results in a negative MC which is shown (open circle) in Fig 8a where the solid line in the figure represents fitting with HLN equation (Eq. 1). We obtain parameters $\alpha$ = -0.07(5) and $l_{\phi}$ = 28.3(2) nm. The obtained value of $\alpha$ is significantly lower than the expected -0.5 for single channel WAL behavior. The decrease of $\alpha$ (still negative) at 2 K is likely due to an effect of magnetic state which strongly reduces the quantum interference effect through breaking time reversal symmetry. The similar decrease of $\alpha$ and a systematic crossover between WAL and WL have been observed with magnetic impurity (Cr) in Bi$_2$Se$_3$ topological insulator (TI) system ($\alpha$ decreases from -0.40 to 0.0 with $\sim$7\% of Cr).\cite{liu} Similar quenching of WAL has been observed in other TI Bi$_2$Te$_3$ doped with magnetic Fe impurity, showing a crossover from WAL to $B^2$ dependance of MC data with higher concentration of Fe.\cite{he} In case of topological surface states, theoretical study suggests an opening of energy gap at Dirac point with magnetic impurity which causes such crossover from WAL to WL.\cite{lu} A transition from WAL to WL has further been discussed in case of graphene, showing it is caused by an intervalley scattering which reduces the chiral nature of Dirac fermions.\cite{cann,tikh} As evident, the situation in TI and graphene are different than the iridium oxides which are primarily dominated by SOC, but the root cause remains similar which requires a certain type of symmetry breaking. The present $l_{\phi}$ is lower than the values obtained for TI systems that are in range of 100 - 500 nm. This low value of $l_{\phi}$ can arise due to an increased magnetic scattering from ordered moments. This impurity induced low values of $\alpha$ and $l_{\phi}$ are also observed in case of other iridates. For instance, the film of Na$_2$IrO$_3$ shows $\alpha$ and $l_{\phi}$ in the range of -0.034 to -0.20 and 27 to 11 nm, respectively in the temperature range of 2 to 25 K.\cite{jender} Nonetheless, an interplay between SOC and magnetic moment on WAL in bulk 5$d$ oxide is quite intriguing. Here, it worth to mention that WL/WAL has originally been discussed at the background of conducting materials but recently this analysis has been tested in semiconducting iridate system.\cite{jender} Iridates with reasonable SOC and exotic band properties, indeed offer an ideal playground to test theoretical models having topological significances. 

\begin{figure}
	\centerline{
		\includegraphics[width=8cm]{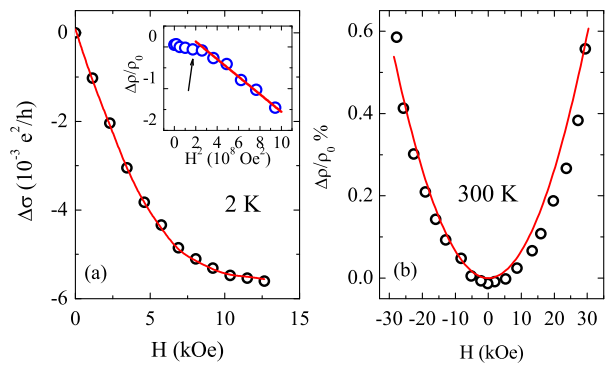}}
\caption{(a) Fitting of magnetoconductivity data with HLN equation (Eq. 1) is shown for Sr$_2$IrO$_4$ at 2 K. Inset shows quadratic field dependance of MR data during initial application of field at 2 K where the straight line is due to fitting of straight line. (b) MR data at 300 K (open circle) and the solid line is due to fitting with quadratic field dependance for Sr$_2$IrO$_4$. The MR data is multiplied by 100 to increase is clarity.}
	\label{fig:Fig8}
\end{figure}

In Fig. 8b, the MR of Sr$_2$IrO$_4$ at 300 K follows parabolic field dependance. However, no dip in MR is observed close to origin which may be due to an effect of high temperature. The dip in MR is sensitive to many parameters. In thin film of Bi$_2$Te$_3$, the dip feature in MR is shown to largely depend on the angle between film plane and magnetic field and magnetic impurities.\cite{he} However, the MR behavior at 300 K can be attributed to Lorentz deflection of charge carriers. The solid line in Fig. 8b is due to fitting with Kohler's formula, $\Delta \rho/\rho(0)$ = $(\mu B)^2$ which gives charge carrier mobility $\mu$ around 250 cm$^2$ V$^{-1}$ s$^{-1}$ which is in the range of semiconducting systems. 

The Fig. 7 further shows that the negative MR during first application of field increases almost linearly above $H_c$, except for parent Sr$_2$IrO$_4$. The nonlinear negative MR at 2 K for Sr$_2$IrO$_4$ follows a $B^2$ dependence in high field regime (see inset of Fig. 8b). In low field regime, however, a deviation from $B^2$ dependence is observed which marks a crossover from initial positive to negative MR with increasing magnetic field (Fig. 7a). This $B^2$ dependance of MR implies a prominent role of magnetic scattering on electron transport. The influence of magnetic scattering in inducing $B^2$ dependance of MC data has been observed in doped TI Bi$_2$Te$_3$.\cite{he} In doped samples, this $B^2$ dependance is softened, which is probably due to weakening of moment with $x$.   
    
\subsection{Conclusion}
In conclusion, the structural, magnetic and electronic transport properties are studied in series of 5$d$ oxide (Sr$_{1-x}$Y$_x$)$_2$IrO$_4$ with $x$ $\leq$ 0.1. While the system retains its original structural symmetry, the unit cell parameters modify with Y substitution. The spectroscopy reveal that doped Y$^{3+}$ converts equivalent amount of Ir$^{4+}$ into Ir$^{3+}$. Opposed to other electron (La) doped system, the low temperature magnetic and insulating state in Sr$_2$IrO$_4$ is very weakly influenced by Y doping. This implies that generated Ir$^{3+}$ ions less likely act for site dilution, instead participate in exchange interaction with Ir$^{4+}$. All samples are highly insulating where the mode of charge conduction follows Mott's 2D VRH model. The MR at low temperature (2 K) magnetic state shows large hysteresis and a sign change with field while in room temperature PM state, MR exhibits only a positive value. This sign change in MR is believed to be caused by an interplay between SOC and magnetic moment. Further, the positive MR at 2 K is explained with WAL behavior. The present results suggest that the conventional picture of $J_{eff}$ under strong SOC needs to be revisited while the alternative pictures of SOC for non-cubic octahedral environment are already being considered. At the same time, local structural as well as magnetic investigations need to be initiated to comprehend this complex magnetism in these materials.

\section{Acknowledgment} 
We acknowledge IISER, Pune and Sunil Nair for the magnetization measurements. We acknowledge DST-FIST and DST-PURSE for funding the `low temperature high magnetic field' and `helium liquefier' facility, respectively. INB acknowledges CSIR, India for fellowship.

\end{document}